\documentclass[A4,twocolumn]{article} 
\usepackage{graphicx}
\usepackage{psfrag}
\usepackage[OT1]{fontenc}
\usepackage[latin1]{inputenc} 
\usepackage[english]{babel}
\usepackage{amssymb}
\usepackage{amsmath}
\begin{document}
\title{Monodisperse Dry Granular Flows on Inclined Planes : Role of Roughness}
\author{Celine GOUJON, Nathalie THOMAS and Blanche DALLOZ-DUBRUJEAUD\\
\\
IUSTI-CNRS UMR 6595, Polytech'Marseille, Technop\^ole de Ch\^ateau-Gombert, \\
13453 Marseille, France.}
\date{}
\maketitle
\abstract{
Recent studies have pointed out the importance of the basal friction on the dynamics of granular flows. We present experimental results on the influence of the roughness of the inclined plane on the dynamics of a monodisperse dry granular flow. We found experimentally that it exists a maximum of the friction for a given relative roughness. This maximum is shown to be independent of the angle of the slope. This behavior is observed for four planes with different bump sizes (given by the size of the beads glued on the plane) from 200 $\mu m$ to 2 $mm$. The relative roughness corresponding to the maximum of the friction can be predicted with a geometrical model of stability of one single bead on the plane. The main parameters are the size of the bumps and the size of the flowing beads. In order to obtain a higher precision, the model also takes into account of the spacing between the bumps of the rough plane. Experimental results and model are in good agreement for all the planes we studied. Other parameters, like the sphericity of the beads, or irregularities in the thickness of the layer of glued particles, are shown to be of influence on the friction.
%
		}
%

		\section{Introduction}

The flow of dense granular matter on inclined planes is often encountered in engineering applications involving the transport of materials such as minerals and cereals. It is also common in geophysical situations where rock avalanches, landslides and pyroclastic and debris flows are natural events consisting in large-scale flows of grains \cite{Takahashi,Campbell,Naaim}. Many chute flow experiments have been carried out and different configurations have been used, changing the boundary conditions from smooth \cite{Patton,Ahn} to rough \cite{Augenstein}, and using different kinds of materials \cite{Ridgway,Robinson,Savage,Drake} without any systematic investigation.

Literature on gravitational dense granular flows on inclined planes shows three regimes of motion depending on the angle of the slope. For small angles, the flow is decelerated; for intermediate angles, the flow reaches a constant mean velocity; and for large angles of inclination, the flow is constantly accelerated \cite{Forterre}. Recent studies have pointed out the importance of the basal friction on the dynamics of granular flow. They provide a model for slow frictional flows based on averaging equations (S$^{\rm t}$ Venant equations)\cite{Pouliquen}. This model has been succesfully tested on non-stationnary flows \cite{Forterre} and is very promising for the dynamics of granular flow. Nevertheless, one of the main parameter of the model is the basal friction. This friction is defined by the tangential stress at the base divided by the normal stress at the base of the flow. In this paper, the word ``friction" refers then to this global bulk effective friction of the flowing granular material down the slope and not to the local contact friction between two beads. This friction represents the interaction between the rough plane and the flowing layer. In this paper, we will not measure this friction but interpret experimental results in term of increase or decrease of the friction. To our knowledge, no study has varied the roughness systematically in order to relate basal friction and geometrical characteristics of the substratum. In fact, most of these studies were made with flowing beads of the same size \cite{Canu} or sometimes greater \cite{Pouliquen} than the beads glued on the plane. It can be intuited that the most important parameter for understanding these flows will be the ratio between the size of the flowing beads and the size of the bumps (glued beads). Now, most industrial or natural flows do not propagate on a roughness composed of the same size of particles. Moreover, these flows are often polydisperse in size. For example, geophysical events involve complex heterogeneous granular material made of particles of various size ranging from micrometers to meters, and often mixed with fluids. In such polydispersed material, it is well known that segregation occurs and  that the large particles rise up to the free surface and go forth to the front \cite{Drahun,Felix,Robinson}. So, due to the phenomenon of segregation, the particles in contact with the substratum are different during the flow, involving different relative roughnesses.

It also appears interesting to compare the motion of a granular mass down a rough inclined plane with the motion of a single bead down a rough plane. Many authors studied the trajectories of a single particle moving down a rough surface. Unfortunately, most studies are limited to the case of a single particle greater than the diameter of the glued beads. It has been found that the trajectories of the grains are very dependent on their size ratio, defined by the diameter of the flowing bead divided by the diameter of the glued beads. In this case too, it points out the fact that the relative roughness matters for the motion. A relative large grain has a greater probability to reach the bottom of an inclined surface than a relative small grain. Experimental \cite{Bideau,Riguidel,Jullien,Henrique}, theoretical \cite{Jullien,Henrique,Batrouni,Bocquet,Vasconcelos,Ancey} and numerical \cite{Riguidel,Henrique,Silbert,Silbert2,Dippel,Dippel2,Dippel3} results show that the same three regimes of motion than those existing for granular flows (decelerated flow, flow with a constant mean velocity, and accelerated flow) can be observed for one single bead. All these regimes depend mainly on the inclination of the plane and on the relative size of the bead. The geometry of the rough plane (random spacing \cite{Batrouni,Dippel3}, regular spacing \cite{Dippel3}) and the coefficient of restitution \cite{Dippel,Dippel3} play also a very important role on the motion of the particle. First, the angle for which there is a transition between the decelerated regime and the steady-state regime decreases when the size ratio increases \cite{Dippel3}. Second, in the decelerated regime, the characteristic length of trapping increases with the size ratio and with the angle of inclination \cite{Henrique}. Third, in the regime where the bead reaches a constant velocity, this velocity increases with the size ratio, and with the angle of inclination \cite{Batrouni,Dippel,Dippel3}. These results appear very interesting for modelling and understanding monodisperse flows down a rough inclined plane by considering one single bead on a plane. 

In this paper, we study several systems, constitued of different sizes of beads flowing on different planes, on which one layer of beads with a given diameter has been glued. It appears from these experiments that the relative roughness, defined as the ratio between the size of flowing beads and the size of glued beads is the main parameter. In order to study the influence of the relative roughness on a large range of value, two sets of experiments are performed: 
\begin{itemize}
\item{varying the size of flowing beads on a given plane}
\item{varying the size of the beads glued onto the plane}
\end{itemize}

This study allows to determine for a large range of value (0.1 to 10) the influence of the relative roughness of a plane on the dynamics of a monodisperse dry granular flow. 

The three types of experiments and the measurement methods are presented in part 2. In part 3, we show experimentally the existence of a minimum of the spreading of a mass of granular matter, corresponding to a particular relative roughness. This minimum corresponds also to a minimum of the velocity of the flow and to a maximum of the thickness of the deposit.  In part 4, we study the dependence of this extremum with few parameters. A model of stability predicts this extremum in part 5. The influence of some other experimental parameters is given in part 6.\\

		\section{Experimental configurations}

				\subsection{Experimental device}

\begin{figure*}
\centering
\includegraphics[scale=0.8]{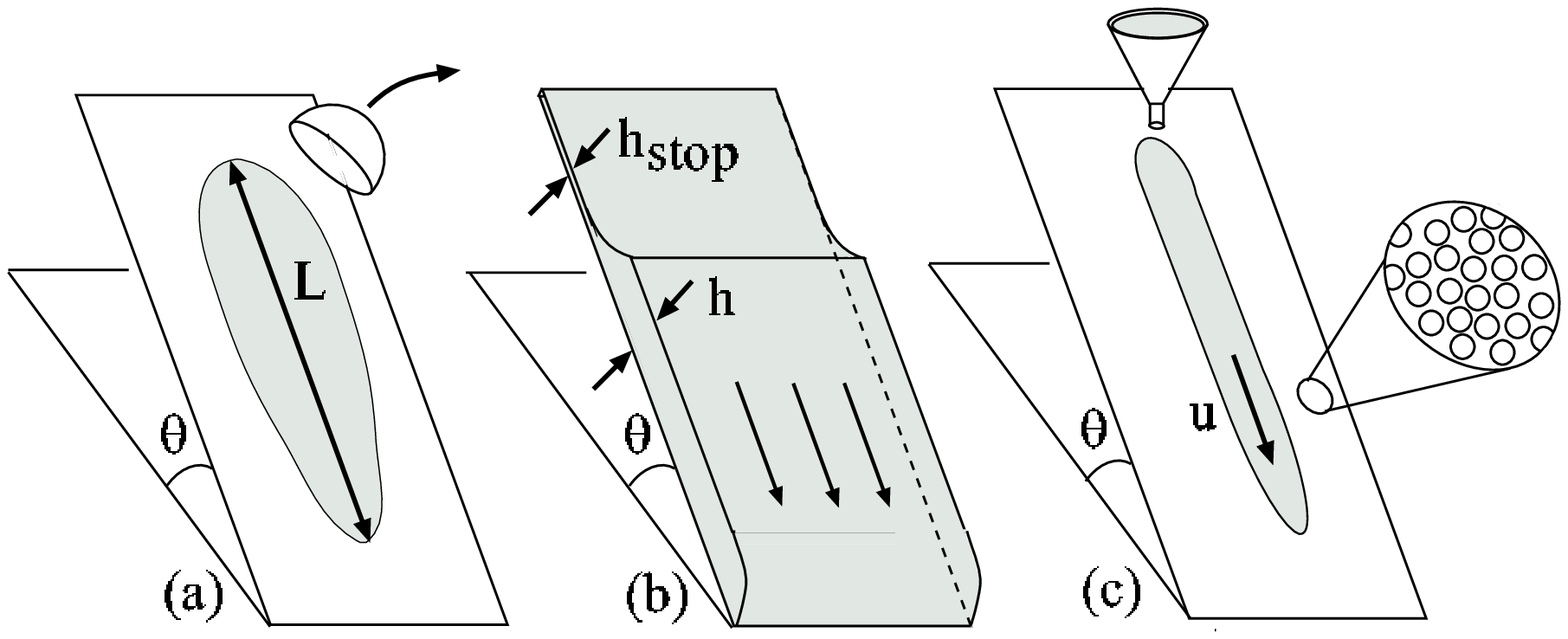}
\caption{Experimental set-up for granular flows on a rough plane: (a) Release of a volume contained in a cap; (b) Deposit left by a uniform steady state flow; (c) Free-side steady state flow. The plane is made rough by gluing beads on it. \label{experience}}
\end{figure*}

The experimental device (fig. \ref{experience}) consists in a 2 m long and 60 cm wide rough plane whose inclination ($\theta$) can be controlled precisely (less than 0.1° of error). The particles are glass beads. We sieve the beads to obtain different ranges between 150 $\mu m$ and 2mm, each range of beads is considered as monodisperse, the width of the distribution being controlled by image analysis. These beads are used for flows and for rough planes (tab. \ref{beads}). We avoid cohesive effect by using beads which have a size greater than 150 $\mu m $. A criterion of sphericity is taken as $d_{max}/d_{min}\leq 1.1$, where $d_{max}$ and $d_{min}$ are the two diameters of the elliptic projection of the particle. The percentage of these non-spherical beads is less than 5 per cent in all batches of beads (tab. \ref{beads}). The plane is made rough by gluing one layer of particles onto the plane. For each plane (tab. \ref{plane}), we measure the compactness ($c$) which is equal to the ratio between the projected area occupied by all the beads and the total area. The compactness is an average on fifteen measurements each taken on a rectangle of a few hundred $mm^{2}$ area and a mean spacing is deduced. Error is given by the standard deviation of the fifteen measurements. We can also directly determine from the images of the plane, a value for the mean spacing which is the mean of all the spacings between the beads of the plane. With these results, we calculate the standard deviation obtained from the distribution of the spacing. The two values of the mean spacing are very close (table \ref{plane}). 

\begin{table*}
\centering
\begin{tabular}{|c|c||c|c|}
\hline
mean size ($\mu m$) & range ($\mu m$) & mean size ($\mu m$) & range ($\mu m$)\\
\hline
150 & 140 - 160 & 580 & 560 - 600 \\
180 & 160 - 200 &  615 & 600 - 630\\
225 & 200 - 250  & 655 & 610 - 700\\
275 & 250 - 300 & 670 & 630 - 710\\
300 & 290 - 310 & 780 & 710 - 850\\
327 & 300 - 355 & 925 & 850 - 1000\\
377 & 355 - 400 & 1400 & 1250 - 1550\\
425 & 400 - 450 & 1850 & 1700 - 2000\\
475 & 450 - 500 & 2000 & 1900 - 2100\\
530 & 500 - 560 & 2150 & 2000 - 2300\\
 & & 5000 & 4900 - 5100\\
\hline
\end{tabular}
\caption {Mean sizes of beads and corresponding ranges of sizes\label{beads}}
\end{table*}

\begin{table*}
\centering
\begin{tabular}{|c|c|c|c|c|c|}
\hline
plane number & 1 & 2 & 3 & 4 \\
\hline
diameter of glued beads ($\lambda$) & 225 $\mu m$ & 425 $\mu m$ & 655 $\mu m$ & 2 mm\\
mean spacing ($\epsilon$) & 46 $\mu m$& 121 $\mu m$ & 160 $\mu m$ & 393 $\mu m$\\
standard deviation of the mean spacing & 41 $\mu m$ & 79 $\mu m$ & 129 $\mu m$ & 376 $\mu m$\\
compactness ($c$) & 0.63$\pm$0.03 & 0.55$\pm$0.04 & 0.57$\pm$0.03 & 0.65$\pm$0.05\\
spacing deduced from $c$ ($\epsilon$) & 44 $\mu m$ & 120 $\mu m$ & 171 $\mu m$ & 362 $\mu m$\\
\hline
\end{tabular}
\caption{Rough planes \label{plane}}
\end{table*}

On each plane, three types of experiments are carried out (fig. \ref{experience}). The first set of experiments consists in the instantaneous release of a fixed volume of beads placed on the plane. The volume is placed in a hemisphere which is instantaneously removed. The released material flows down the slope, spreads and ultimately stops leaving a tear-shaped deposit, characterized by its length $L$ and its width $W$. In order to control precisely the amount of the material present in the cap, the mass of beads poured into the cap is weighed before each run, and is equal to 176 g. In the second set of experiments, we study, for each range of beads and on each plane, the evolution of the thickness of the deposit ($h_{stop}$), left by a wide uniform steady state flow, for several inclinations of the slope ($\theta$). Third, we study, the velocity ($u$) of a stationary non uniform flow, produced by a constant flux input. Subsequently, we define the following parameters :
\begin{itemize}
\item $d$ : the diameter of flowing beads 
\item $\lambda$: the diameter of beads glued on the plane
\item $c$ : the compactness of the layer of beads glued on the plane
\item $\epsilon$ : the mean gap between two beads of the plane calculated from the compactness assuming a triangular lattice $ \epsilon = \left(\large{\sqrt{\frac{\pi}{2c\sqrt{3}}}}-1\right)\lambda$
\item $\theta$ : the angle of inclination
\end{itemize}

				\subsection{Measurement methods}

The thickness of the deposit is measured precisely by the deviation of the projection of a laser sheet. Two different pictures are compared: the first one is obtained with no beads on the plane (that is $h_{stop}=0$), the second one with the deposit of a uniform steady flow. The deviation is proportional to its thickness ($h_{stop}$). The scale is determined by the deviation given by a plate whose thickness is known precisely. With this method, thickness is measured with less than 100 $\mu m$ of error.

For the measurements of the velocity of the non-uniform steady state flow, some marks are drawn along the plane, every ten centimeters. It allows us to measure the velocity on a video and also to verify if the flow is in a steady state. 

		\section{Existence of a maximum of the friction}
			\subsection{Results for the plane number 2}
In this part, all the experiments are carried out on the plane number 2. By keeping the same plane, and by varying the size of flowing beads, the flowing beads are submitted to a variable relative roughness. In fact, one plane can be rough for some beads, and smooth for others. The couple (plane, flowing beads) allows us to define the relative roughness of our system. Another mean to vary this relative roughness would be to vary the size of glued beads and not the size of flowing beads, but it is less precise because it is difficult to control the spacing between the glued beads with accuracy.

\begin{figure}
\includegraphics[width=7.8cm]{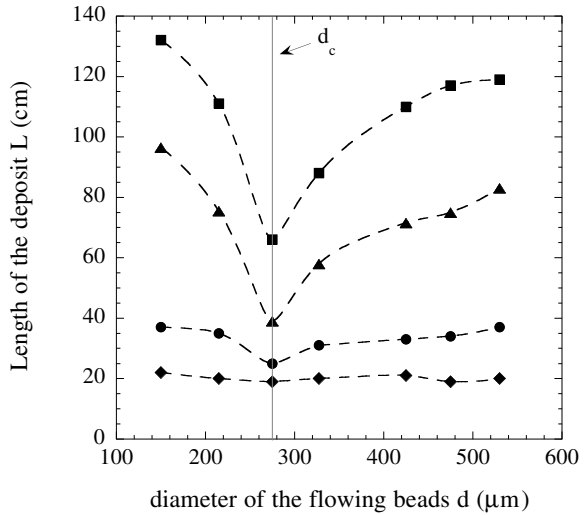}
\caption{Length of the deposit as a function of $d$ for different angles of inclination: ($\blacksquare$) $\theta$ =28,3°; ($\blacktriangle$) $\theta$=25,7°; ($\bullet$) $\theta$=22,8°; ($\blacklozenge$) $\theta$=18,4°. There is a diameter $d_{c}$, for which the length presents a minimum. $d_{c}$ is independent of  $\theta$ (plane 2).\label{L}}
\end{figure}

First, the experiments of the release of a constant volume are done to study the length of spreading ($L$) versus the diameter of  beads flowing ($d$). These experiments are carried out for different inclinations ($\theta$). Results are presented on figure \ref{L}. For an angle of inclination greater than 28°, the measured lengths are underestimated for beads which have a size greater than  327 $\mu m$ (425 $\mu m$, 475$\mu m$, 530 $\mu m$), because some beads do not stay on the plane. For these sizes, the percentage of loosing beads is respectively equal to 3, 8,  13 \%. For an angle of inclination smaller than 28°,  all the beads stay on the plane. The length presents a minimum for a given size of beads, named $d_{c}$ (fig. \ref{L}). This diameter is independent of the angle of inclination. But, the minimum of the length of the spreading is more enhanced for large angles of inclination. 
\\
\begin{figure}
\includegraphics[width=7.8cm]{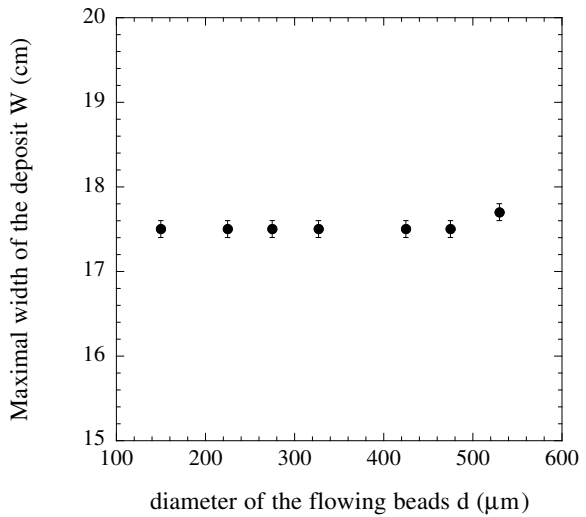}
\caption{The maximal width of the deposit is independent of $d$ ($\theta$=25,7°, plane 2).\label{W}}
\end{figure}

On the previous experiments, we also measure the maximal width ($W$) of the deposit, for different sizes of beads (fig. \ref{W}). The maximal width is independent of the size of the beads flowing. 
 \\

\begin{figure}
\includegraphics[width=7.8cm]{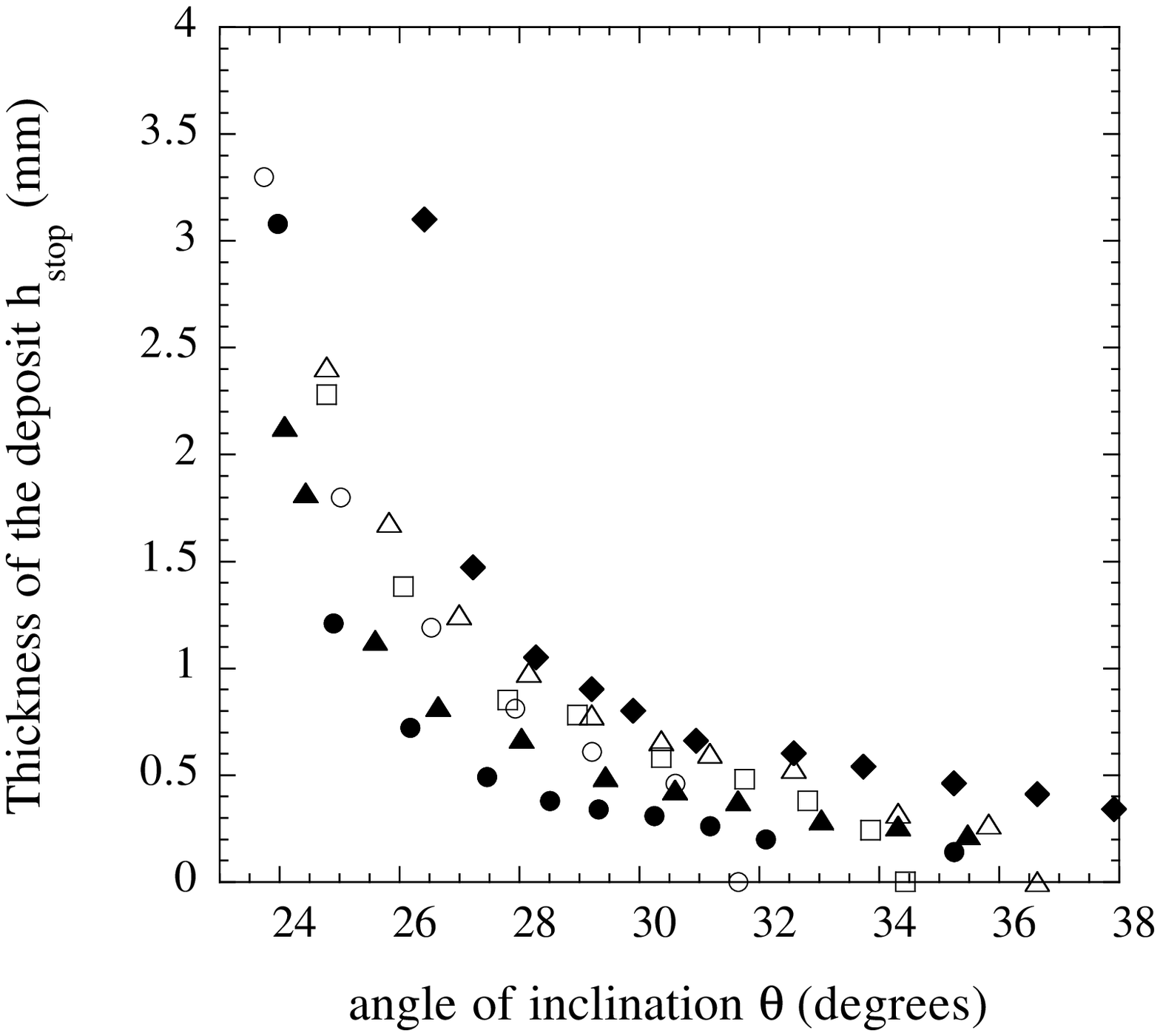}
\caption{$h_{stop}$ for different sizes of beads on plane 2: ($\bullet$) $d$=150 $\mu m$; ($\blacktriangle$) $d$=225 $\mu m$; ($\blacklozenge$) $d$=275 $\mu m$; ($\vartriangle$) $d$=327 $\mu m$; ($\square$) $d$=425 $\mu m$; ($\circ$) $d$=530 $\mu m$. For $d=d_{c}$, the $h_{stop}$ curve is maximum ($d_{c}=275 \mu m$). \label{hstopteta}} 
\end{figure}

\begin{figure}
\includegraphics[width=7.8cm]{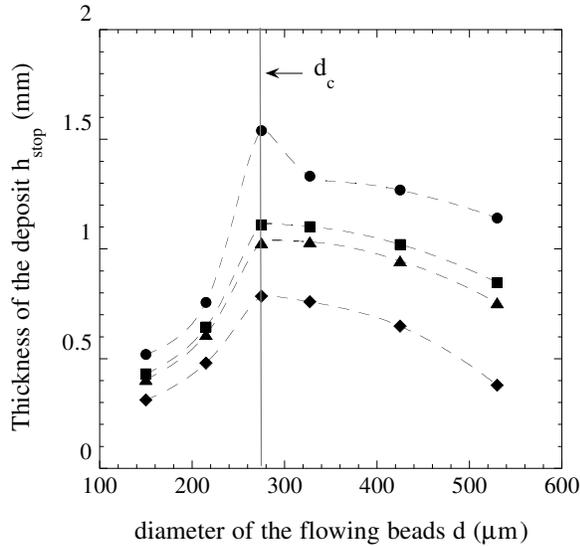}
\caption{$h_{stop}$ for different angles of inclination: ($\bullet$) $\theta$=27°; ($\blacksquare$) $\theta$=28°; ($\blacktriangle$) $\theta$=28.3°; ($\blacklozenge$) $\theta$=30°. For $d=d_{c}$, $h_{stop}$ is maximum (plane 2). \label{hstopvsd}} 
\end{figure}

Second, experiments of wide flows give the ability to study the thickness of the deposit ($h_{stop}$) left by a uniform steady state flow. For these experiments, the inclination and the size of flowing beads vary. Figure \ref{hstopteta} presents the curves of $h_{stop}$ as a function of $\theta$. The curves are classified, all over the range of inclinations we study, and $h_{stop}$ is greater for $d=d_{c}$ than for other sizes. Figure \ref{hstopvsd} presents $h_{stop}$ as a function of $d$ for different angles of inclination. For $d=d_{c}$, $h_{stop}$ is greater than for other sizes, and that for every angle of inclination. 
\\

\begin{figure}[!h]
\includegraphics[width=7.8cm]{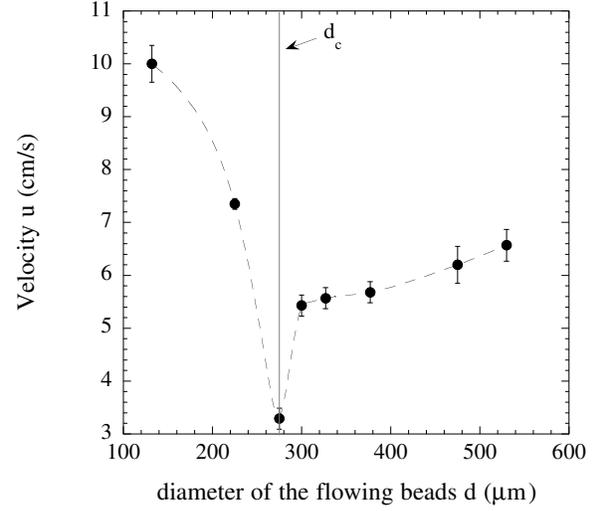}
\caption{ Velocity as a function of $d$. For $d=d_{c}$, it exists a minimum of the velocity ($\theta$=25.7°, plane 2). \label{v}}
\end{figure}

Third, velocity ($u$) is measured on steady state free-side flows (fig. \ref{v}). As in the previous experiments, the behavior of the flow presents a singularity for $d=d_{c}$ : the velocity is minimum for this value. 
\\
\begin{figure}
\includegraphics[width=7.8cm]{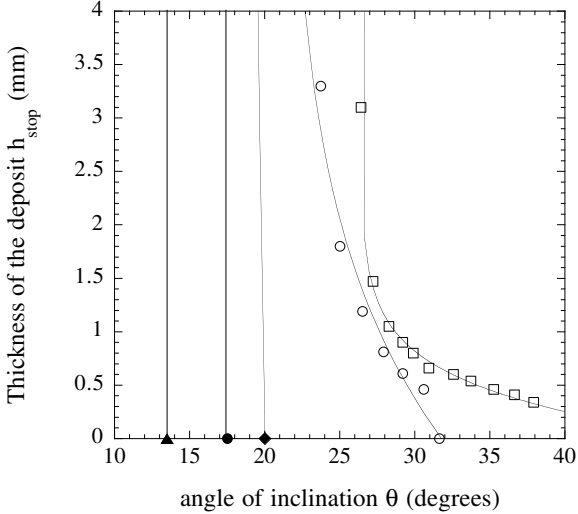}
\caption{$h_{stop}$ for different sizes of large flowing beads: ($\square$) $d$=275 $\mu m$; ($\circ$) $d$=530 $\mu m$; ($\blacklozenge$) $d$=1 mm; ($\bullet$) $d$=2 mm; ($\blacktriangle$) $d$=5 mm (plane 2). The curvature of $h_{stop}$ tends to 0 and $\theta_{1}$ decreases for large beads. \label{hstopgdd}}
\end{figure}
\\
We notice that the curves L, $h_{stop}$ and $u$ as a function of the diameter of the flowing beads $d$ are not symmetrical in relation to $d=d_{c}$.
In these three types of experiments, we point out that it exists a value of the diameter of flowing beads, named $d_{c}$ for which :
\begin{itemize}
\item the thickness of the deposit ($h_{stop}$) is maximum
\item the length of the deposit ($L$) is minimum
\item the velocity of the non-uniform steady state flow ($u$) is minimum
\end{itemize}
The experiments show that the diameter $d_{c}$ is independent of the angle of inclination of the slope $\theta$. These results can be linked to the basal friction of the flow. For $d<d_{c}$, the friction increases with the diameter of beads flowing. For $d>d_{c}$, the friction decreases when the diameter of beads flowing increases, and there is a maximum of this friction for $d=d_{c}$. 

To found the same extremum on these 3 parameters could be predictable, because the three parameters, $L$, $h_{stop}$ and $u$ are linked. We approximate the surface of the deposit by a rectangle and assume that the volume ($V$) is equal to: 
\\
\begin{equation}
V=W L h_{stop}
\end{equation}
and as the volume of flowing beads is constant, deduce that $L$ and $h_{stop}$ vary as the inverse of each other because $W$ is constant (fig. \ref{W}). On the other hand, if we suppose that results given by Pouliquen \cite{Pouliquen} are verified in our configuration, the velocity of a steady state flow is a function of  $h_{stop}$ and of the height of the flow ($h$):
\begin{equation}
\frac{u}{\sqrt{gh}}=\beta\frac{h}{h_{stop}}
\end{equation}
with $\beta$ a constant. A maximum for $h_{stop}$ at $d=d_{c}$ is then equivalent to a minimum of the velocity $u$ at the same value of $d$.  

We could interpret the existence of the minimum of the spreading for the two borderline cases: $d\gg\lambda$ and  $d\ll\lambda$.

For $d\gg\lambda$, the length $L$ and the velocity $u$ increase with $d$ and with the angle of inclination $\theta$ (fig \ref{L}, \ref{v}). In fact, beads are greater than $\lambda$ and than the spacing between the glued beads of the plane ($\epsilon$), the relative roughness is small and beads flow easily. Flows are submitted to decreasing friction when $d$ increases. This behavior is in accordance with the results of previous studies on the motion of a large single bead whose velocity \cite{Riguidel,Vasconcelos,Dippel} or characteristic length of trapping \cite{Henrique} increases with the size ratio and with the angle of inclination $\theta$. The range of angles of inclination [$\theta_{1};\theta_{2}$] (for $\theta=\theta_{1}$, $h_{stop}$ $\rightarrow$ $\infty$, for $\theta=\theta_{2}$, $h_{stop}$=0) for which a steady state flow happens, decreases when d increases . For greatest $d$ (fig \ref{hstopgdd}), this range of angles of inclination is very small ($d$=1 $mm$) and tends to 0 ($d$=2 $mm$, $d$=5 $mm$). At the same time, both $\theta_{1}$ and $\theta_{2}$ decrease. There is a limit for which we can only define one angle. For an angle greater than this angle, no deposit stays on the plane. For an angle of inclination lower than this angle, no steady state flow is observed. When $d$ increases, the value of this limit angle decreases, beads flow easily. The global shift of $h_{stop}$ curve toward small angles when d increases could be related to the decrease of  the angle separating steady motion from the decelerating motion of one single bead on a rough plane \cite{Dippel3}, although, the values of this angle are very small (approximatively 9° for a size ratio equal to 1) compared to $\theta_{1}$ and $\theta_{2}$.

For $d\ll\lambda$, the length $L$ increases when $d$ decreases, which is more surprising. An explanation could be that the holes filled with small beads reduce the roughness. 

But this interpretation of the two borderline cases do not allow us to explain all our results because the evolution of the length with the diameter seems continuous and does not only show a transition below a critical size which fills holes. Consequently the variations of the friction with $d$ is not due only to the filling or not of the holes of the plane. We will discuss more precisely this point in part 4.

		\subsection{Results for different planes}

\begin{figure}[!h]
\includegraphics[width=7.8cm]{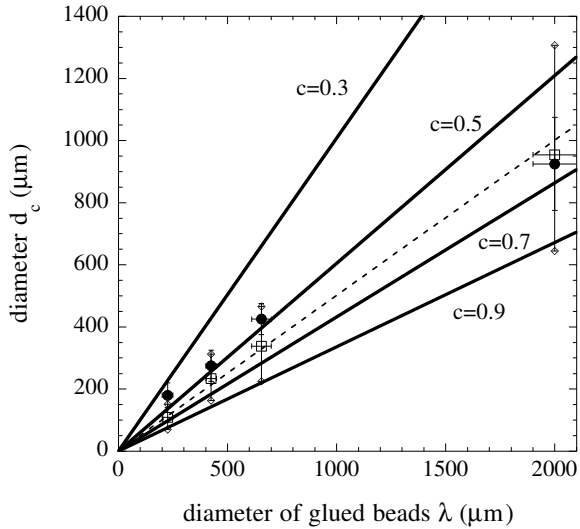}
\caption{$d_{c}$ obtained on different planes are close to $\lambda$/$2$ (dotted line): ($\bullet$) experimental results; ($\square$) model with the experimental parameters values. Curves represent predictions of the model taking into account of different compactnesses ($c$). The model predicts with accuracy the value of $d_{c}$. For the model, error bars ($\lozenge$)come from the fact that c and $\lambda$ are not well-determined. For the experiments, $d_{c}$ is not known precisely because of the discontinuity of the values of $d$ (-). The error bars (axis $\lambda$) for the model and the experiments are the same, taking into account of the range of sizes of the glued beads. \label{dcass}}
\end{figure}

In this part, we study the behavior of beads on the different planes, presented in part 1. We measure for each plane and for different ranges of beads, the length $L$ of the spreading of a constant mass and $h_{stop}$. For all these planes, there is a diameter of beads which presents a minimum of length $L$ and a maximum of $h_{stop}$. We compare the diameter of these beads ($d_{c}$) with the diameter of beads glued on our plane ($\lambda$) (fig. \ref{dcass}).

The beads which correspond to the minimum of length are not always the same. This minimum is then not an experimental artefact due to some characteristics of a batch of beads. 

$d_{c}$ is approximatively equal to $\lambda$/2 with little deviations from this value. But, in all of our experiments, the variation of $d$ is discontinuous, and we can not determine with accuracy the value of $d_{c}$ for which the friction is really maximum.
\\
\\
	\section{Implications on the roughness of filling the holes}

\begin{figure}[!h]
\includegraphics[width=7.8cm]{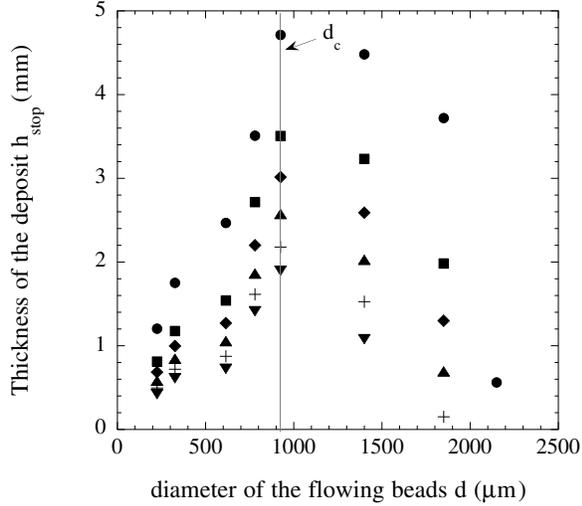}
\caption {$h_{stop}$ on the plane number 4 (2mm glued beads) for different angles of inclination: ($\bullet$) $\theta$=24°;  ($\blacksquare$) $\theta$=26°; ($\blacklozenge$) $\theta$=27°; ($\blacktriangle$) $\theta$=28°; (+) $\theta$=29°; ($\blacktriangledown$) $\theta$=30°. There is a diameter $d_{c}$=925 $\mu m$ for which $h_{stop}$ is maximum. \label{hstop2000}} 
\end{figure}

It appears also interesting to understand the impact on the friction of filling the holes of the plane by small beads. We used the plane number 4 made by gluing 2mm beads, so it is possible to study a wider range of small $d$ without using cohesive small beads less than 150 $\mu m$. Small beads, for which the ratio $d / \lambda$ is very small, fill the holes of the plane. For example, for 225 $\mu m$ beads, one hole of the plane contains approximatively 30 beads. The measurements of $h_{stop}$ and $L$ are carried out on plane number 4 and only some results of $h_{stop}$ are presented on figure \ref{hstop2000}. The behavior is similar to the one observed previously, with $h_{stop}$ maximum and $L$ minimum for $d_{c}$=925 $\mu m$.

\begin{figure}[!h]
\includegraphics[width=7.8cm]{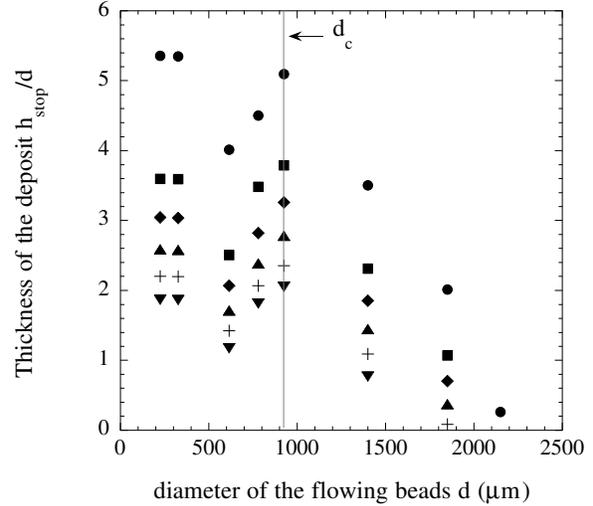}
\caption {$h_{stop}/d$ on the plane number 4 (2mm glued beads) for different angles of inclination: ($\bullet$) $\theta$=24°;  ($\blacksquare$) $\theta$=26°; ($\blacklozenge$) $\theta$=27°; ($\blacktriangle$) $\theta$=28°; (+) $\theta$=29°; ($\blacktriangledown$) $\theta$=30°. There is a diameter $d_{c}$=925 $\mu m$ for which $h_{stop}/d$ is locally maximum. \label{hstop2000d}} 
\end{figure}

If we compare the thickness of the deposit in term of number of beads ($h_{stop}/d$), we observe that (fig. \ref{hstop2000d}):
\begin{itemize}
\item{$d_{c}$ could be a local extremum (the friction could be greater for the smallest beads)}
\item{the curves corresponding to $d$=225 $\mu m$ and $d$=327 $\mu m$ are the same}.
\end{itemize}

For the smallest beads, the value of $h_{stop}/d$ seems to tend to an asymptotic value.

This behavior was not observed for the other planes, because the ranges of the flowing beads compare to the glued one, were not wide enough.				
\\
\\

For the smaller beads, all the holes of the plane are filled, and the basal plane is a mixture of large beads (2mm) and small beads (225 $\mu m$ or 327 $\mu m$). One possibility to interpret the fact that $h_{stop}$/$d$ for $d$=225 $\mu m$ and for $d$=327 $\mu m$ are equal is to consider that the friction comes principally from the filled holes and that the contribution of the summit of large glued beads to the friction is negligible. We can suppose that the beads flow on a plane which has a roughness given by the size of beads organized with a 3D random compactness in the holes. In that case, the behavior is independent of the size of the beads explaining the similarity between the curves for 225 $\mu m$ and 327 $\mu m$.

Because of the filling of the holes, we also suppose that 225 $\mu m$ flowing on 2 mm rough plane could be similar to 225 $\mu m$ flowing on 225 $\mu m$ rough plane. It is so interesting to compare 225 $\mu m$ beads flowing on :
\begin{itemize}
\item{a 225 $\mu m$ plane ($c=0.63$), plane 1}
\item{a 2mm plane ($c=0.65$), plane 4}
\end{itemize}

In the second case, the beads flow on a plane constituted of 225 $\mu m$ not glued small beads and very smooth rounded surfaces (top of 2 mm beads). Inside the holes, the lattice of 225 $\mu m$ beads is randomly distributed with a compactness between 0.5 and 0.7. This compactness is also the compactness of a plane section of a random three dimensional piling. 

\begin{figure}[!h]
\includegraphics[width=7.8cm]{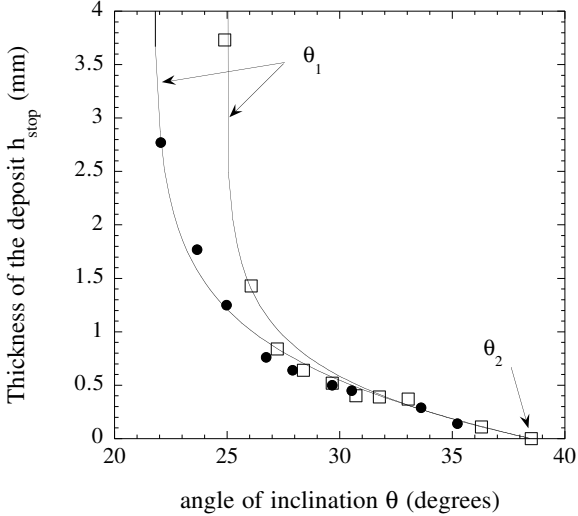}
\caption { $h_{stop}$ as a function of  $\theta$ for the same 225 $\mu m$ beads flowing, on different planes: ($\bullet$) plane number 1 ($\lambda=225 \mu m$); ($\square$) plane number 4 ($\lambda=2mm$). The curves represent the data of $h_{stop}$ fitted by the function (\ref{fit}) and the two associated parameters $\theta_{1}$ and $\theta_{2}$. {\label{h200}}}
\end{figure}

The $h_{stop}$ curves are very close each other except at small angles (fig. \ref{h200}). From these results, we assume that 225 $\mu m$ beads flowing on a 2 mm plane could be equivalent to 225 $\mu m$ beads flowing on a 225 $\mu m$ plane which have a compactness equal to the compactness of a random lattice. This result is surprising because it means that there is almost no influence of the smooth part of large beads tops, at least when the deposit is thin. For small angles of inclination, the difference could be explain by the difference in the values of the compactness. 

Notice that for 225 $\mu m$ beads flowing on a 2mm plane, filling the holes corresponds to introduce rough surface between the large beads. But for 2mm flowing beads, filling the holes with 225 $\mu m$ corresponds to a smoothing of the plane.
\\
\\
\\
		\section{Model of stability for the existence of the maximum}
				\subsection{Stability of a single bead on a rough plane }

We have seen in the previous part, that the parameters, $h_{stop}$, $L$ and $u$, all present an extremum for the same value of $d_{c}$. We consider, in this model, a rough plane defined by two parameters: the diameter of beads glued $\lambda$, and the mean spacing between two beads glued on the plane $\epsilon$. The model consists in determining the angle corresponding to the limit of stability of one single bead put on the plane. The criterion of choice of the exact location of this bead is such that the fall of this particular bead put $h_{stop}$ equal to 0. Among all the possible locations, we choose the more stable one, which would correspond to the last bead to fall before $h_{stop}=0$ in the case of a whole layer of beads covering the plane. According to this criterion, we calculate an angle corresponding to the limit where there would be no more deposit in the configuration of our experiments. 

This angle can be compared to another angle, $\theta_{2}$ obtained from the $h_{stop}$ versus $\theta$ experimental curves. $\theta_{2}$ is defined by the angle for which $h_{stop}$($\theta$)=0 and would then correspond to our criterion previously explained. In fact, from each curve $h_{stop}$ as a function of  $\theta$, we can deduce the two parameters, $\theta_{1}$ and $\theta_{2}$ from the curves (fig. \ref{h200}). For $\theta \leq \theta_{1}$, no uniform steady flow can be observed ($h_{stop}=\infty$). For $ \theta \geq  \theta_{2} $ no deposit stays on the plane ($h_{stop}=0$) \cite{Pouliquen}. We can also determine more precisely the values of $\theta_{2}$ by using the whole set of experimental data of $h_{stop}$ fitted by the following function \cite{Pouliquen}:
\begin{equation}
tan\theta=tan\theta_{1}+\left(tan\theta_{2}-tan\theta_{1}\right)exp\left(-\frac{h_{stop}}{Ld}\right) 
\label{fit}
\end{equation}

We also experimentally determine that the level $h_{stop}=0$ corresponds to the plane tangential to the top of the beads glued on the plane, because there is no deviation of the laser projection when the holes are filled with very small beads. That means that a plane where only the holes are filled with small beads will be considered as having no deposit on it with our measurement method. From these observations, we can define a criterion for the choice of the bead of which we will consider the stability: the bead chosen is the lowest bead which top is above the level $h_{stop}$=0. Consequently, if the bead is small enough to fit between the glued beads of the plane, the lowest bead is the bead, which top is tangential to the level $h_{stop}$=0. (fig. \ref{criterion})

In our model, we determine the minimum angle for the chosen bead to fall down. This angle is named the angle of stability. Our model is then based on a phenomenon of start because we consider the stability of a bead, which is initially at rest. The angle of stability, we determine, is the angle for which this bead becomes unstable. However, $\theta_{2}$ is experimentally the angle for which no deposit stays on the plane after a flow. Consequently, the measurements of $h_{stop}$ are determined by a phenomenon of stop. But neglecting the effect of the inerty in the process of stop, the angle of stability is then equal to $\theta_{2}$. 
\\

We first explain our model of stability in a two-dimen\-sional case, the same arguments being valid in the three-dimensional case. The results presented in the figures (\ref{dcass}, \ref{t2}) are from the three-dimensional case calculations. 
\\
			\subsubsection{The two-dimensional case}

\begin{figure}[!h]
\includegraphics[scale=0.35]{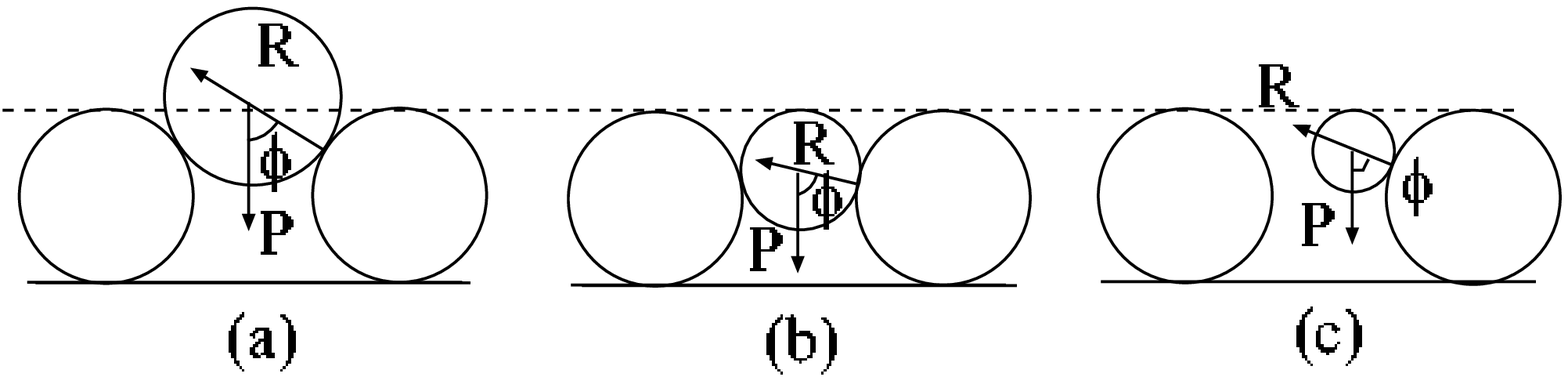}
\caption{Criterion of stability in different cases: (a) $d > d_{h}$; (b) $d=d_{h}$ and (c) $d<d_{h}$. The dotted line represents the level $h_{stop}=0$. The stability of the bead is maximum for $d=d_{h}$. \label{criterion}}
\end{figure}

The bead which stability we consider is tangential to one glued bead of the rough plane, and its top is above or tangential to the level $h_{stop}$=0. We define the diameter ($d_{h}$), which is the diameter of the bead tangential to the two beads of the rough plane, and tangential to the level $h_{stop}$=0 (fig. \ref{criterion} case b). For beads greater than $d_{h}$, the chosen bead is above the level $h_{stop}$=0, and tangential to the two beads of the plane (fig. \ref{criterion} case a). For beads smaller than $d_{h}$, the chosen bead, is tangential to the level $h_{stop}$=0 and to one bead of the rough plane, the holes are filled with small beads (fig. \ref{criterion} case c).

Once the location of the bead chosen, the criterion of stability is very simple. The bead is submitted to 2 forces, its weight, named $P$, and the contact force between the bead and one bead of the plane, named $R$ (fig. \ref{criterion}). The direction of  $R$ depends on the angle of inclination of the plane $\theta$. We consider the angle between these two forces, $\phi$. If this angle $ \phi$ is equal to 0, the bead is unstable, else the bead is stable. 

For the cases presented in figure \ref{criterion}, we calculate the angle of inclination for which the bead, initially at rest, is unbalanced. This angle is equal to $\theta_{2}$ because for an angle lower than the angle of stability of this bead, $h_{stop}$ is not equal to 0, and for an angle of inclination greater than this angle of stability, the bead falls leading to $h_{stop}$=0. 
\\

With this choice of the unstable bead, and with this criterion of stability, it appears that the angle of stability is maximum for the bead whose centre is the lowest; i.e. for the bead with the diameter $d_{h}$. We realize with this schema that the maximum of the friction comes from a pure geometrical reason.
\\
		\subsubsection{The three-dimensional case}

\begin{figure}[!h]
\centering
\includegraphics[scale=0.4]{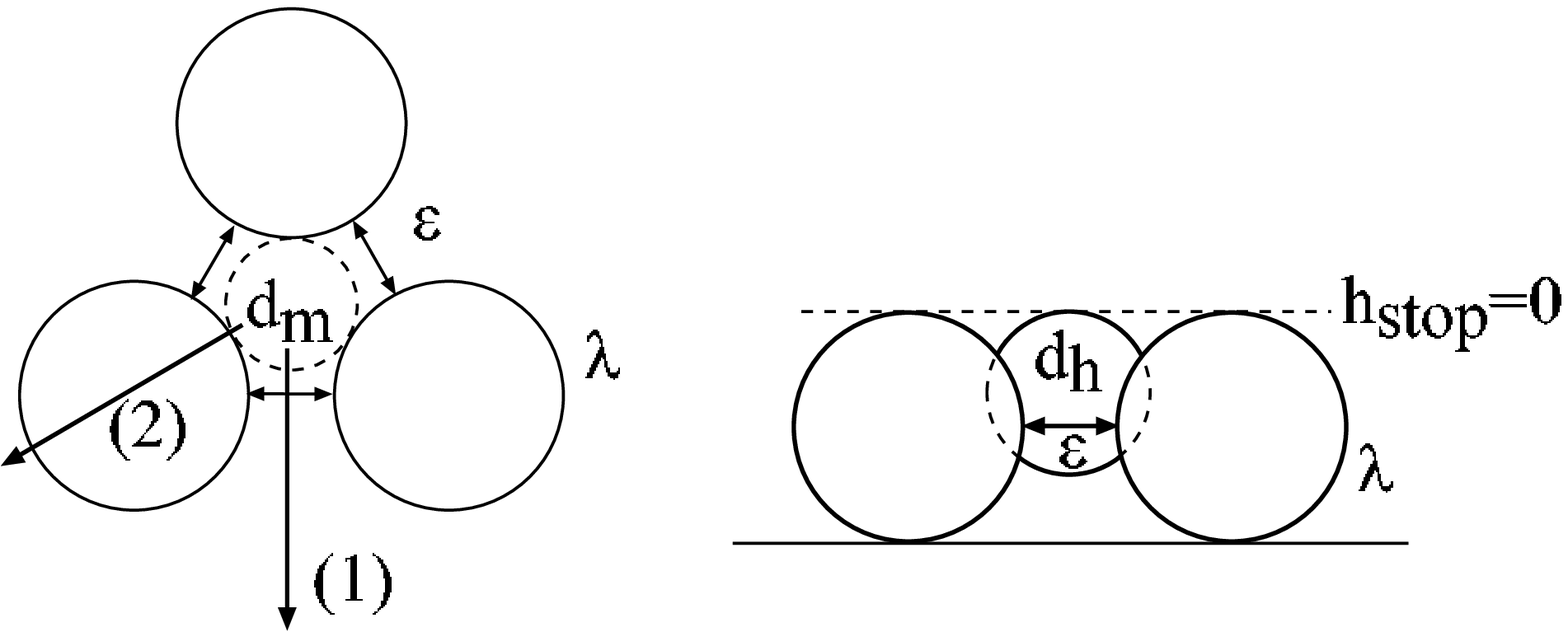}
\caption{Representation of the roughness plane triangular lattice, with a spacing $\epsilon$. $d_{m}$ is the diameter maximum of the bead that can be placed in between the three beads of the rough plane. The direction (1) is for the bead which pass over the gap between two beads of the rough plane, the direction (2) is for the bead which get over one glued bead. $d_{h}$ is the diameter of the bead tangential to the three glued beads and to the level $h_{stop}$=0. \label{frugueux}}
\end{figure}

In that section, we consider the stability of the bead which is tangential to 2 beads of the rough plane, and which top is still above or tangential to the level $h_{stop}$=0. We assume that the plane is ordered with a triangular lattice (fig. \ref{frugueux}). 
We calculate the diameter maximum ($d_{m}$) of a bead, lodging between the three beads glued onto the plane. We also define the diameter ($d_{h}$) of the bead which is tangential to the three beads of the rough plane and to the level $h_{stop}=0$. This diameter is equivalent to $d_{h}$ defined in the two-dimensional case, but their values are not the same.

The criterion of stability is the same, but we could determine a whole range of angles of stability. The smallest, is for the bead which passes over the gap between two beads of the rough plane (fig. \ref{frugueux}; direction 1), the highest is for the bead which gets over one bead of the rough plane (fig. \ref{frugueux}; direction 2). In the following, we only consider the smallest angle of stability (fig. \ref{frugueux}; direction 1). 

The choice of the bead is separated in several cases: for $d \geq d_{h}$, the case is well-defined: the chosen bead is tangential to the three beads of the rough plane. The calculations, for the case $d\geq \lambda$, have been made by Riguidel \cite{Jullien}. Here, we obtain the same results for $d \geq d_{h}$. For  $d<d_{h}$, we study two cases. For the largest $d<d_{h}$, the chosen bead is tangential to two beads of the rough plane and to the level $h_{stop}$=0. For the smallest $d$, the bead is too small to be tangential to two beads and to the level $h_{stop}$=0. In that case, the beads, which are also smaller than $d_{m}$, fill the holes between the large glued beads. Because there is a  filling of the holes, we are not sure that the more unstable bead is one touching a large glued bead, or one being on the surface of the filled hole. At the limit $d\rightarrow$0, we expect that there is no more effect of the top of the large glued beads on the stability of the chosen small beads. We choose to consider the stability of a small bead which stays on the random loose packing of the others small beads. For that reason, the angle of stability would not vary with the size of the small beads for $d\rightarrow$0. For the calculations involving the smallest beads, we thus consider the stability of a small bead tangential to three small beads and to the level $h_{stop}$=0 (being at the top of the random piling which fills the holes); the three small beads spacing being determined by the assumption of a random packing in the holes between the large glued beads. In all our calculations, we choose the compactness of the surface of a random packing equal to 0.57. We discuss this value of the compactness in part 5.2. The diameter for which, we pass from one calculation to another is not well defined in our model. Consequently for $d<d_{h}$, the angle of stability will be taken equal to the maximum of the 2 angles given by the two cases. 

The curve is decreasing for large beads and increasing for small beads (fig. \ref{t2}). For small beads, the intersection of the two calculations of $\theta_{2}$ happens for $d$ very close to $d_{m}$, with $d_{m}$:
\begin{equation}
d_{m}=\frac{2\left(\lambda+\epsilon \right)}{\sqrt{3}}-\lambda
\end{equation}
Consequently the stability of beads smaller than $d_{m}$ is the same wathever their size are.

The main result is that the angle of stability $\theta_{2}$ presents a maximum for $d_{h}$:
\begin{equation}
d_{h}=\frac{\left(\lambda+\epsilon\right)^{2}}{3\lambda}=\frac{\pi}{6c\sqrt{3}}\lambda
\label{dh}
\end{equation}
One more time, we notice that the maximum is due to a pure geometrical reason. The model predicts that $d_{h}$ is proportional to $\lambda$ but also that it depends on the spacing of the glued beads $\epsilon$.
\\

We have also calculated the angle of stability, considering the possible patterns fitting in the holes according to the size ratio between flowing and glued beads. But it appears that the results are not really different except that there are some discontinuities when passing from a pattern to another. The global behavior being the same, we choose not to represent the results.
\\
			\subsection{Comparison between the experimental results and results obtained from the model}

\begin{figure}[!h]
\includegraphics[width=7.8cm]{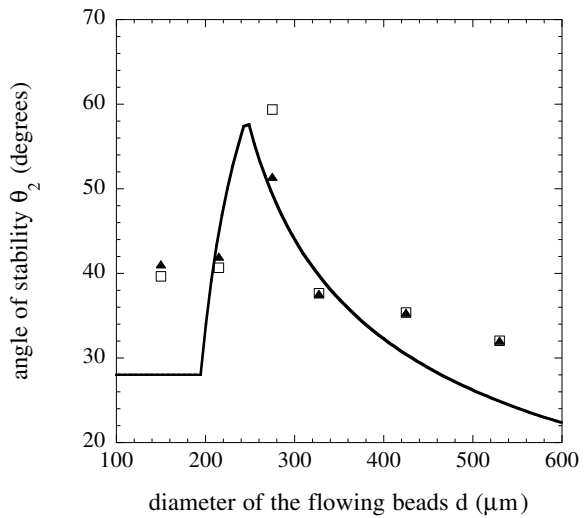}
\caption{$\theta_{2}$ as  a function of d. ($\blacktriangle$) experimental results obtained on the plane number 2 ($\lambda$=425 $\mu m$, c=0.55) ($\Box$) experimental results obtained by the fit (\ref{fit}). The curve represents the results given by the model with $\lambda$=425 $\mu m$ and $c=0.55$ \label{t2}}
\end{figure}

In figure \ref{t2},  we compare, for one plane, experimental results, and results obtained from the model. We have also compared the results for the others planes; the same following remarks being valid. The global behavior of experimental results and model are the same: for $d<d_{m}$ $\theta_{2}$ is constant, for $d_{m}<d <d_{h}$, the angle $\theta_{2}$ increases with $d$, and for $d >d_{h}$, $ \theta_{2}$ decreases. The curvature for experimental results and results predicted by the model seems to be the same. Taking into account of the discontinuity in the experimental variation of $d$, the model predicts with accuracy the value of the extremum, and in the following text, $d_{c}$ is put equal to $d_{h}$. In contrast, for small diameter of the flowing beads, the value of $\theta_{2}$ given by the model is underestimate. In fact, we have seen that the criterion of stability is very difficult to apply precisely  for the small beads. The compactness of the top of the random piling we have choosen is equal to 0.57. However, all we know is that the value of the compactness is between 0.3 and 0.9 (see the end of this section), which respectively correspond to an angle of stability $\theta_{2}$ equal to 90 ¡ and 19,7 ¡. This range of values of the compactness is not precise but we do not consider valuable to calculate accurately the fraction because the top of the random piling is probably not equivalent to a flat pattern of small glued beads with a certain spacing. Nevertheless, the model shows that it exists a maximum and predicts with accuracy the value of $d$ corresponding to the maximum (fig. \ref{dcass}).

In figure \ref{dcass},  $d_{h}$ obtained from the model for all planes, are plotted as a function of the diameter of glued beads ($\lambda$). The model predicts that $d_{h}$ is proportionnal to  $\lambda$, when no variations of the compactness. But the compactness changes from one plane to another explaining that the values of $d_{c}$ are not aligned. We expected that all the experimental points will be between the two lines corresponding to compactness 0.5 and 0.7, in accordance with the measured compactnesses of the plane (tab. 2) which is approximatively verified.

Errors bars for results from the model (fig. \ref{dcass}) take into account of the uncertainty standard deviations on the values of the compactness because the measurements could be not representative, the surface analysis being only equal to few thousand $mm^2$. Moreover, we suppose in the model that only the mean value of the compactness matters and not the variability of the spacing around its mean value. Errors bars for the model take into account of the standard deviation of the direct measurements of the spacing, and suggest that this variability is important. Nevertheless, assuming values for $\lambda$ and $\epsilon$, the model predicts to our satisfaction the value of $d_{c}$. However, molecular dynamic simulations on the the motion of a single particle on a bumpy inclined plane \cite{Dippel3,Batrouni} show that intoducing disorder has a similar influence on the motion of the ball as the introduction of a regular spacing. The velocity of the ball is not affected by the introduction of disorder to the line, the velocity can be approximated by the mean velocity on a line with equally spaced balls: this spacing corresponds to the mean value of the disordered case. 

Because experimental reasons, all the planes, we made, have also a compactness between 2 values: $c_{min}$ and $c_{max}$. $c_{max}$ is the compactness for which all the beads glued are in contact. This compactness is equal to $c_{max}={\pi}/{2\sqrt{3}}=0.9$ for monodisperse size beads. The minimum of the compactness is determined by the manufacturing process. To make the planes, we put a large amount of beads on a sticky paper, we reduce the spacing between the beads by pressing on the beads and remove the beads not glued. No hole larger than a bead size stays empty. The minimum of the compactness is obtained when the spacing between three beads ($d_{m}$) is equal to $\lambda$, leading to a compactness equal to $c_{min}$=0.3. With these extreme compactnesses, the model predicts that the value of the diameter $d_{c}$ is between these two lines, and consequently, $d_{c}$ is always less than  $\lambda$ (fig. \ref{dcass}).
\\
\\
		\section{Influence of others parameters}

\begin{figure}[!h]
\includegraphics[width=7.8cm]{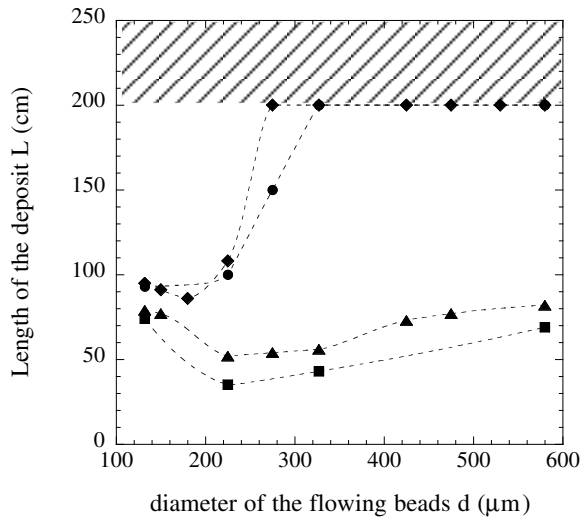}
\caption{ Length as a function of $d$: ($\blacklozenge$) spherical beads flowing on a one-layer spherical beads plane (case 1); ($\bullet$) non-spherical beads flowing on a one-layer spherical beads plane (case 2); ($\blacktriangle$) spherical beads flowing on a multi-layer plane (case 3); ($\blacksquare$) non-spherical beads flowing on a multi-layer plane (case 4). $L=200cm$ represents experiments where no deposit stays on the plane ($\lambda$=225 $\mu m$ $\theta$=25.7°). \label{sphere}}
\end{figure}

In the previous part, we worked with rough planes made of one layer of spherical beads. It appears also interesting to study the influence of the quality of beads and planes. First, we will study the influence of the sphericity of glued and flowing beads. ``non-spherical'' beads batch is composed to approximatively of 60\% of spherical beads, of 15\% of splinters and of 25\% of non-spherical beads, with the criterion of sphericity defined in part 2. Second, we will study the influence of a mutli-layered rough plane whose amount of layer is spacially variable. In figure \ref{sphere}, we study four different configurations for the length of the spreading ($L$) as a function of the diameter of the flowing beads $d$:
\begin{itemize}
\item{(1) spherical beads flowing on the plane number 1 on which one layer of spherical beads are glued, named ``good'' plane}
\item{(2) non-spherical beads flowing on the ``good'' plane}
\item{(3) spherical beads flowing on a plane on which there are, depending on the location, approximatively one to three layers of 225 $\mu m$ non-spherical beads, named ``bad'' plane}
\item{(4) non-spherical beads flowing on the ``bad'' plane.}
\end{itemize}

				\subsection{Influence of sphericity of beads}

In this part, we compare the experimental results obtained with the configurations 1 and 2 (or 3 and 4). In each case, spherical and non spherical beads flow on the same plane. We notice that it exists, in these four cases, a diameter for which the length of spreading is minimum. The lengths obtained on the same plane are very similar although the lengths for non spherical beads are 1.6 smaller than for spherical beads in the cases 3 and 4. The sphericity of the flowing beads seems to have just a little influence on the spreading. In fact, during the flow, because of the segregation, the splinters, which represents 15\% of the volume, segregate at the surface and at the front, and they consequently are probably of little influence on the flow basal friction.

				\subsection{Influence of a multilayered rough plane}

In this part, we compare the experimental results obtained with the configurations 1 and 3. Spherical beads flow on a ``good'' plane and on a ``bad'' plane. We see that the quality of the plane is of great importance. If we compare the configurations 2 and 4, done with ``non-spherical'' beads, we see the same effect. 

First, the length of spreading is greater in the case of a ``good'' plane than in the case of a ``bad'' plane. This result may be explain by the fact that the presence of splinters increases the roughness of the ``bad'' plane. Moreover, the steps between the layers create a barrier difficult to get over. That is why the lengths of flowing are really smaller in the case of a ``bad'' plane. 

Second, we observe, that in the case of a ``good'' plane, for a size of beads greater than  a limit size (275 $\mu m$ for these experiments), no deposit stays on the plane (length is arbitrary marked as 200 cm in that case). But in the case of a ``bad'' plane, there is a deposit for larger flowing sizes. The behavior of the beads, in the case of they are greater than $d_{c}$ is totally different. This behavior can be also interpreted as the ``bad'' plane being more rough than the ``good'' one.

We notice that it exists always a diameter $d_{c}$ for which the length is minimum. Unfortunately, our data are not precise enough to conclude if the value $d_{c}$ is the same in the two cases. 
\\

In conclusion, the sphericity of flowing beads has less influence, the global behavior is not affected by the quality of the bead. But the quality of the plane is of great importance, the main effect is that the relative roughness of a plane can be totally smooth (cases 1 and 2) or still rough (cases 3 and 4) for large flowing beads.

		\section{Conclusion}

This paper presents experiments of flows of various sizes of glass beads down rough inclined planes. Varying the size appears equivalent to a variation of the relative roughness of the plane. We show the existence of a size of beads ($d_{c}$) for which the length of the spreading and the velocity are minimum and the thickness of the deposit is maximum; these facts being interpreted as a maximum of the friction. This behavior is robust, being observed for several planes of different roughnesses. We also study the influence of different parameters on the value of the diameter $d_{c}$ corresponding to the maximum of relative roughness. First, it is independent of the angle of inclination. Second, it is mainly determined by the size  $\lambda$ of the particles glued on the plane, but the compactness also has an important influence on it. When the compactness of the rough plane increases, the value of $d_{c}$ decreases. A simple geometrical model of stability of one bead allows to estimate the value of this diameter $d_{c}$, by calculating the value of the angle of inclination above which no deposit stays on the plane.

Most flows are more complex than the simple avalanche of monodispersed glass beads. The interactions between the plane and the flow will be difficult to understand because each type of particle experiences a different relative roughness. For example, in the case of a bidisperse granular flow, the existence of a minimum for $d_{c}$ leads probably to complex behaviors depending on the positions of each sizes compared to $d_{c}$ combined to the segregation effect. Our results provide a good frame for the interpretation of the dynamics of the flow of a granular polydisperse matter down a rough plane.

	\section{Acknowledgments}

This research was supported by the French Ministry of Research and \'Education. We thank O. Pouliquen and Y. Forterre for fruitful discussions and F. Ratouchniak for his technical assistance.


\begin{thebibliography}{99}
\bibitem[1]{Takahashi}T. Takahashi, Debris Flows, Annu. Rev. Fluid Mech. \textbf{13}, (1981) 57 
\bibitem[2]{Campbell}C.S. Campbell, P.W. Cleary and M. Hopkins, Large-scale landslide simulations: Global deformations, velocities and basal friction, J.  Geophys.  Res. \textbf{100}, (1995) 8267 
\bibitem[3]{Naaim} M. Naaim, S. Vial and R. Couture, Saint-Venant approach for rock avalanches modelling, Presses de l'Ecole Nationale des Ponts et Chaussées, Paris, (1997)
\bibitem[4]{Patton} J.S. Patton, C.E. Brennen and R.H. Sabersky, Shear flows of rapidly flowing granular materials, ASME, J. Appl. Mech. \textbf{52}, (1987) 172 
\bibitem[5]{Ahn} H. Ahn, C.E. Brennen and R. H. Sabersky, Measurements of velocity, velocity fluctuation, density and stresses in chute flows of granular materials, ASME, J. Appl. Mech. \textbf{59}, (1991) 119 
\bibitem[6]{Augenstein} D.A. Augenstein and R. Hogg, Friction factors for powder flow, Powder Technology \textbf{10}, (1974) 43 
\bibitem[7]{Ridgway}K. Ridgway and R. Rupp, Flow of granular material down chutes, Chemical and Process Engineering, (1970) 82-85 
\bibitem[8]{Robinson}D.A. Robinson and S.P. Friedman, Observations of the effects of particle shape and particle size distribution on avalanching of granular media, Physica A \textbf{311}, (2002) 97-110 
\bibitem[9]{Savage}S.B. Savage, Gravity flow of cohesionless granular materials in chutes and channels, J. Fluid Mech. \textbf{92}, (1979) 53-96  
\bibitem[10]{Drake}T.G. Drake, Granular flow: Physical experiments and their implications for microstructural theories, J. Fluid Mech. \textbf{225}, (1991) 121-152 
\bibitem[11]{Forterre}O. Pouliquen and Y. Forterre, Friction law for dense granular flows:application to the motion of a mass down a rough inclined plane J. Fluid. Mech. \textbf{453}, (2002) 133-151 
\bibitem[12]{Pouliquen}O. Pouliquen, Scaling laws in granular flows down rough inclined planes, Physics of Fluids \textbf{11}, (1999) 542-548 
\bibitem[13]{Canu}A.C. Santomaso and P. Canu, Transition to movement in granular chute flows, Chemical Engineering Science \textbf{56}, (2001) 3563-3573 
\bibitem[14]{Drahun} J.A. Drahun and J. Bridgwater, The mechanisms of free surface segregation, Powder Technology \textbf{36}, (1983) 39-53 
\bibitem[15]{Felix} G. Felix and N. Thomas, Relation between granular flow regimes and morphology of the deposits: formation of levées in pyroclastic deposit, Earth Planetary Sciences Letters (submitted).
\bibitem[16]{Bideau}M.A. Aguirre, I. Ippolito, A. Calvo, C. Henrique and D. Bideau, Effects of geometry on the characteristics of the motion of a particle rolling down a rough surface, Powder Technology \textbf{92}, (1997) 75-80 
\bibitem[17]{Riguidel}F.X. Riguidel, A. Hansen and D. Bideau, Gravity-Driven motion of a particle on an inclined plane with controlled roughness, Europhys.  Lett. \textbf{28}, (1994) 13 
\bibitem[18]{Jullien}F.X. Riguidel, R. Jullien, G. Ristow, A. Hansen and D. Bideau, Behaviour of a sphere on a rough inclined plane J.Phys. I \textbf{4}, (1994) 261 
\bibitem[19]{Henrique}C. Henrique, M.A. Aguirre, A. Calvo, I. Ippolito, S. Dippel, G.G. Batrouni and D. Bideau, Energy dissipation and trapping of particles moving on a rough surface, Phys. Rev. E \textbf{57}, (1998) 4743-4750 
\bibitem[20]{Batrouni}G.G. Batrouni, S. Dippel and L. Samson, Stochastic model for the motion of a particle on an inclined rough plane and the onset of viscous friction, Phys. Rev. E \textbf{53}, (1996) 066496-066503 
\bibitem[21]{Bocquet}L. Bocquet, J. Errami and T.C. Lubensky, Hydrodynamic Model for a Dynamical Jammed-to-Flowing Transition in Gravity Driven Granular Media, P.R.L. \textbf{89}, (2002) 184301 
\bibitem[22]{Vasconcelos}G.L. Vasconcelos and J.J.P. Veerman, Geometrical model for a particle on a rough inclined surface, Phys. Rev. E \textbf{59}, (1999) 5641-5646  
\bibitem[23]{Ancey} C.Ancey, P.Evesque and P.Coussot, Motion of a single bead on a bead row: theoretical investigations, J.Phys. I \textbf{6}, (1996) 725-751 
\bibitem[24]{Silbert}L.E. Silbert, D. Ertas, G.S. Grest, T.C. Halsey, D. Levine and S.V. Plimpton, Granular flow down an inclined plane: Bagnold scaling and rheology Phys. Rev. E \textbf{64}, (2001) 051302 
\bibitem[25]{Silbert2}L.E. Silbert, J.W. Landry, G.S. Grest, Granular flow down a rough inclined plane: Transition between thin and thick piles, Physics of Fluids \textbf{15}, (2003) 1-10  
\bibitem[26]{Dippel}S. Dippel, G.G. Batrouni and D.E. Wolf, How transversal fluctuations affect the friction of a particle on a rough inclined plane, Phys. Rev. E \textbf{56}, (1997) 3645-3656 
\bibitem[27]{Dippel2}S. Dippel, D.E. Wolf, Molecular Dynamics simulations of granular chute flow, Computer Physics Communications \textbf{121}, (1999) 284-289 
\bibitem[28]{Dippel3}S. Dippel, G.G. Batrouni and D.E. Wolf, Collision-induced friction in the motion of a single particle on a bumpy inclined line, Phys. Rev. E \textbf{54}, (1996) 6845-6856 
\end{thebibliography}
\end{document}